\magnification=1200
\baselineskip=13pt
\overfullrule=0pt
\tolerance=100000
\nopagenumbers

\font\tenbifull=cmmib10 \skewchar\tenbifull='177
\font\tenbimed=cmmib7   \skewchar\tenbimed='177
\font\tenbismall=cmmib5  \skewchar\tenbismall='177
\textfont9=\tenbifull
\scriptfont9=\tenbimed
\scriptscriptfont9=\tenbismall

\mathchardef\alpha="710B
\mathchardef\beta="710C
\mathchardef\gamma="710D
\mathchardef\delta="710E
\mathchardef\epsilon="710F
\mathchardef\zeta="7110
\mathchardef\eta="7111
\mathchardef\theta="7112
\mathchardef\iota="7113
\mathchardef\kappa="7114
\mathchardef\lambda="7115
\mathchardef\mu="7116
\mathchardef\nu="7117
\mathchardef\micron="716F
\mathchardef\xi="7118
\mathchardef\pi="7119
\mathchardef\rho="711A
\mathchardef\sigma="711B
\mathchardef\tau="711C
\mathchardef\upsilon="711D
\mathchardef\phi="711E
\mathchardef\chi="711F
\mathchardef\psi="7120
\mathchardef\omega="7121
\mathchardef\varepsilon="7122
\mathchardef\vartheta="7123
\mathchardef\varphi="7124
\mathchardef\varrho="7125
\mathchardef\varsigma="7126
\mathchardef\varpi="7127

 at 8truept

\

{\hfill \hbox{\vbox{\settabs 1\columns
\+ solv-int/9706005\cr
}}}
\centerline{}
\bigskip
\bigskip
\bigskip
\baselineskip=18pt

\centerline{\bf A Lax Description for Polytropic Gas Dynamics}
\vfill
{\baselineskip=11pt
\centerline{J. C. Brunelli\footnote{*}{e-mail address: brunelli@fsc.ufsc.br}}
\medskip
\medskip
\centerline{Universidade Federal de Santa Catarina}
\centerline{Departamento de F\'\i sica -- CFM}
\centerline{Campus Universit\'ario -- Trindade}
\centerline{C.P. 476, CEP 88040-900}
\centerline{Florian\'opolis, SC -- BRAZIL}
\medskip
\medskip
\centerline{and}
\medskip
\medskip
\centerline{Ashok Das}
\medskip
\medskip
\centerline{Department of Physics and Astronomy}
\centerline{University of Rochester}
\centerline{Rochester, NY 14627 -- USA}
}
\vfill

\centerline{\bf {Abstract}}

\medskip

We give a Lax description for the system of polytropic gas equations. The 
special structure of the Lax function naturally leads to the two infinite 
sets of conserved charges associated with this system. We obtain closed form 
expressions for the conserved charges as well as the generating functions for 
them. We show how the study of these generating functions can naturally lead 
to the recursion relation between the conserved quantities as well as the 
higher order Hamiltonian structures.
\medskip

\vfill
\eject
\headline={\hfill\folio}
\pageno=1

Systems of hydrodynamic type have been extensively studied 
over the last several years [1-5]. These are first-order quasilinear systems of partial 
differential equations. A sub class of these systems is the two-component 
hyperbolic system of conservation laws [5] which has the general Hamiltonian 
form (in $1+1$ dimensions)
$$
\pmatrix{u\cr 
\noalign{\vskip 0.3 truecm}%
v}_t=
\pmatrix{0&\partial\cr
\noalign{\vskip 0.3 truecm}%
\partial&0}
\pmatrix{{\delta{\cal H}\over\delta u}\cr
\noalign{\vskip 0.3 truecm}%
 {\delta{\cal H}\over\delta v}}\eqno(1)
$$
This can also be written as
$$
\pmatrix{u\cr 
\noalign{\vskip 0.3 truecm}%
v}_t=
\pmatrix{{\partial^2H\over\partial u\partial v}&
{\partial^2H\over\partial v\partial v}\cr
\noalign{\vskip 0.3 truecm}%
{\partial^2H\over\partial u\partial u}&{\partial^2H\over\partial u\partial v}}
\pmatrix{u\cr
\noalign{\vskip 0.3 truecm}%
 v}_x\eqno(2)
$$
where
$$
{\cal H}[u,v]=\int dx\,H(u,v)\eqno(3)
$$
is an Hamiltonian functional of zeroth order. An infinite number of zeroth order conserved charges are associated with this system of equations. In fact, if
$$
{\cal Q}[u,v]=\int dx\, Q(u,v)\eqno(4)
$$
represents a conserved charge of the system, it is easy to show that the 
density must satisfy
$$
{\partial^2 H\over\partial v^2}{\partial^2 Q\over\partial u^2}=
{\partial^2 H\over\partial u^2}{\partial^2 Q\over\partial v^2}\eqno(5)
$$
In particular, we note that  $H$ satisfies this equation.

Of special interest are systems for which (5) admits a separation of variables 
such that
$$
{\partial^2 Q\over\partial u^2}={\lambda(u)\over\mu(v)}
{\partial^2 Q\over\partial v^2}\eqno(6)
$$
A separable Hamiltonian of the form
$$
H(u,v)=-\left({1\over 2}u^2v+F(v)\right)\eqno(7)
$$
describes the physically important system of gas dynamics [6] corresponding to 
$\lambda(u)=1$ and $\mu(v)=F''(v)/v$. There is yet another important system -- the 
nonlinear elastic medium model [6] -- which is also described by a separable 
Hamiltonian of the form
$$
H(u,v)=-\left({1\over 2}u^2+{\widetilde F}(v)\right)\eqno(8)
$$

Olver and Nutku [5], in a beautiful paper, gave a systematic description of 
systems like (1). In particular, they have discussed the conserved charges and the Hamiltonian structures 
of the system of polytropic gas, which corresponds to the choice
$$
F(v)={v^{\gamma}\over\gamma(\gamma-1)},\,\,\,\gamma\not=0,1\eqno(9)
$$
in (7), leading to the equations
$$
\eqalign{
u_t+uu_x+v^{\gamma-2}v_x=&0\cr
v_t+(uv)_x=&0
}\eqno(10)
$$
It was shown in [5] that this system of equations has two infinite sets of conserved charges.

However, a Lax  description for such systems is missing so far.
It is known [1,3] that a Lax description for dispersionless systems such as (1) are best given in terms  of polynomial functions in phase space (We will continue to call these as Lax operators.). However, whatever Lax formalism describes such a system, it must have the peculiar property of yielding two infinite sets of conserved charges.  In Ref. 7 a nonstandard Lax description for this system of equations with $\gamma=2$ was proposed, which is also equivalent to the shallow water wave equation. This system of equations is also known as the irrotational Benney's equation [4] and have been well studied in the literature. However, $\gamma=2$ turns out to be a particularly simple choice in that the two sets of conserved charges become degenerate in this case and hence one does not expect any new structure in the Lax description. In going  beyond $\gamma=2$ to $\gamma=3$, one immediately notices that the system of equations (10) can be written as the $2\times2$ matrix Riemann equation
$$
{\bf U}_t={\bf U}{\bf U}_x\eqno(11)
$$ 
where
$$
{\bf U}=\pmatrix{u&v\cr v&u\cr}\eqno(12)
$$
It is quite straightforward to show that  the dispersionless Lax operator ${\bf L}=p^2{\bf I}+{\bf U}$, describes this system of equations with the two infinite sets of conserved charges obtained from $\hbox{Tr}\,{\bf L}^{2n+1\over2}=
\int dx\,\hbox{Res}\left(\hbox{tr}\,{\bf L}^{2n+1\over2}\right)$ taking  the 
matrix trace ``tr'' both with respect to diagonal and off-diagonal elements of the
matrices (``Res'' is the coefficient of the $p^{-1}$ term). 

For any integer $\gamma$, therefore, either a matrix or a scalar generalization of a Lax representation
for (10) can be pursued. We have found that the scalar Lax operator
$$
L=p^{\gamma-1}+u+{v^{\gamma-1}\over(\gamma-1)^2}p^{-(\gamma-1)}\eqno(13)
$$
with the dispersionless nonstandard representation
$$
{\partial L\over\partial t}={(\gamma-1)\over\gamma}
\left\{\left(L^{\gamma\over\gamma-1}\right)_{\ge1},L\right\}\eqno(14)
$$
where $\{A,B\}={\partial A\over\partial x}{\partial B\over\partial p}-
{\partial B\over\partial x}{\partial A\over\partial p}$, leads to the system of
polytropic gas equations (10) (In fact, the hierarchy of equations is obtained from $\left(L^{n+{1\over\gamma-1}}\right)_{\ge1}$). It is interesting to note that the  Lax operator (13) as well as the dynamical equations (14) naturally reduce to the shallow water system for $\gamma=2$ given in Ref. [7]. However, we would like to emphasize here that the Lax function in (13) is quite rigid in that any deformation of this leads to inconsistent equations. It is also  worth noting here that since the system of equations (10) is known to have two distinct recursion operators [5], we suspect that there may be yet another equivalent Lax description of the system. However, we have not succeeded in finding it.

Conserved charges associated with the system can be obtained from (13) through (These are the only fractional powers with nontrivial residues.)
$$
{\overline{\cal H}}_n=\hbox{Tr}\,L^{n+{\gamma-2\over\gamma-1}},\,\,n=0,1,2,3,\dots\eqno(15)
$$
and the first few densities are easily obtained to be
$$
\eqalign{
{\overline H}_0=& {(\gamma-2)\over(\gamma-1)}u\cr
{\overline H}_1=& {(2\gamma-3)(\gamma-2)\over(\gamma-1)^2}\left({1\over2!}u^2+
{1\over(\gamma-1)(\gamma-2)}v^{\gamma-1}\right)\cr
{\overline H}_2=& {(3\gamma-4)(2\gamma-3)(\gamma-2)\over(\gamma-1)^3}\left({1\over3!}u^3+
{1\over(\gamma-1)(\gamma-2)}uv^{\gamma-1}\right)\cr
&\vdots\cr
}\eqno(16)
$$
In general, we can write
$$
{\overline H}_n=(n+1)!\hbox{C}^{(n+1)(\gamma-1)-1\over(\gamma-1)}_{n+1}H_n
$$
where $H_n$'s correspond to one of the infinite sets of conserved densities obtained in [5]. The conserved densities (16) were obtained by expanding $L^{1\over\gamma-1}$ 
around $p=\infty$. However, the present Lax function has a singularity at $p=0$ and, consequently an alternate expansion of the fractional powers is possible around $p=0$ as well [8] and it gives us a second set of nontrivial conserved charges through
$$
{\overline {\widetilde{\cal H}}}_n=\hbox{Tr}\,L^{n+{1\over\gamma-1}},\,\,n=0,1,2,3,
\dots\eqno(17)
$$
where the first few densities are
$$
\eqalign{
{\overline {\widetilde H}}_0=& (\gamma-1)^{-{2\over\gamma-1}} v\cr
{\overline {\widetilde H}}_1=& (\gamma-1)^{-{2\over\gamma-1}}
{\gamma\over(\gamma-1)}uv\cr
\overline{{\widetilde H}}_2=& (\gamma-1)^{-{2\over\gamma-1}}
{\gamma(2\gamma-1)\over(\gamma-1)^2}\left({1\over2!}u^2v+
{v^\gamma\over\gamma(\gamma-1)}\right)\cr
&\vdots\cr
}\eqno(18)
$$
In general, we can write
$$
{\overline{\widetilde H}}_n={n!\over(\gamma-1)^{2\over\gamma-1}}
\hbox{C}^{n(\gamma-1)-1\over(\gamma-1)}_{n}{\widetilde H}_n
$$
where ${\widetilde H}_n$'s correspond to the second set of conserved densities obtained in [5].
Incidentally, this alternate expansion around $p=0$ also gives us a second 
consistent dispersionless Lax equation
$$
{\partial L\over\partial t}=
\left\{\left(L^{\gamma-2\over\gamma-1}\right)_{\le0},L\right\}\eqno(19)
$$
which leads to the set of equations
$$
\eqalign{
u_t=&-\left(v^{\gamma-2}\over(\gamma-2)^2\right)_x\cr
v_t=&-u_x\cr
}\eqno(20)
$$
With proper rescaling, this can be rewritten also as
$$
\eqalign{
u_t=&-v^{\gamma-3}v_x\cr
v_t=&-u_x\cr
}
$$
We recognize this as the elastic medium equations with the polytropic gas potential ${\widetilde F}(v)={v^{\gamma-1}\over(\gamma-1)(\gamma-2)}$ [5]. And our derivation clarifies the fact that both these systems share the same conserved charges because they are obtained from the same Lax function.

We see that the Lax operator (13) nicely reproduces both sets of
densities $H_n$ and ${\widetilde H}_n$ described by Olver and Nutku [5]. We can
even obtain a closed form expression for these densities from their definition and they have the form
$$
\eqalign{
H_n=&\sum_{m=0}^{\left[{n+1\over2}\right]}
\left(-\prod_{k=0}^m{1\over k(\gamma-1)-1}\right)
{u^{n-2m+1}\over m!(n-2m+1)!}{v^{m(\gamma-1)}\over(\gamma-1)^m}\cr
{\widetilde H}_n=&
\sum_{m=0}^{\left[{n\over2}\right]}
\left(\prod_{k=0}^m{1\over k(\gamma-1)+1}\right)
{u^{n-2m}\over m!(n-2m)!}{v^{m(\gamma-1)+1}\over(\gamma-1)^m}\cr
}\eqno(21)
$$
Since these are obtained from a Lax function, it follows that they are conserved under the flow. However, it is also straightforward to check explicitly from the closed form of these charges in (21) that
$$
{\partial^2 H_n\over\partial v^2}=v^{\gamma-3}
{\partial^2 H_n\over\partial u^2},\,\,\,
{\partial^2 {\widetilde H}_n\over\partial v^2}=v^{\gamma-3}
{\partial^2 {\widetilde H}_n\over\partial u^2}
$$
leading (from (5)) to the fact that they are indeed conserved.

These expressions allow us to write the following generating functions [1,3,4]
for the conserved densities (In fact, they also follow from an analysis of the associated linear equations in the dispersionless limit.)
$$
\eqalign{
\chi=&\lambda^{-{1\over\gamma-1}}\left[\left\{\left({u+\lambda\over2}\right)^2-
{v^{\gamma-1}\over(\gamma-1)^2}\right\}^{1/2}+
{u+\lambda\over2}\right]^{1\over\gamma-1}\cr
{\widetilde \chi}=&\lambda^{\gamma-1}
\left[\left\{\left({u+\lambda\over2}\right)^2-
{v^{\gamma-1}\over(\gamma-1)^2}\right\}^{1/2}-
{u+\lambda\over2}\right]^{1\over\gamma-1}\cr
}\eqno(22)
$$
We note that these generating functions reduce to those of Manin [1] for $\gamma=2$. Furthermore, using (10), it is easy to check that
$$
\eqalign{
{\partial\chi\over\partial t}=&{\partial\chi\over\partial u}u_t+
{\partial\chi\over\partial v}v_t={\partial\ \over\partial x}
\left({\gamma-1\over\gamma}\lambda^{-1}\chi^\gamma-u \chi\right)\cr
{\partial{\widetilde \chi}\over\partial t}=&
{\partial{\widetilde \chi}\over\partial u}u_t+
{\partial{\widetilde \chi}\over\partial v}v_t=-{\partial\ \over\partial x}
\left({\gamma-1\over\gamma}\lambda^{-(\gamma-1)^2}
{\widetilde \chi}^\gamma+u{\widetilde \chi}\right)\cr
}\eqno(23)
$$
This shows that $\chi$ and $\widetilde \chi$ are, indeed, conserved for any value of the parameter $\lambda$ and, consequently,  generate conserved densities. In fact, expanding these for large $\lambda$, it is easy to identify
$$
\eqalign{
\chi=&1+\sum_{n=0}^\infty
{(-1)^n\lambda^{-(n+1)}\over((n+1)(\gamma-1)-1)}(n+1)!\,
\hbox{C}^{(n+1)(\gamma-1)-1\over(\gamma-1)}_{n+1}\,H_n\cr
{\widetilde\chi}=&\sum_{n=0}^\infty
{(-1)^n\lambda^{-n}\over(n(\gamma-1)+1)}{n!\over(\gamma-1)^{2\over\gamma-1}}\,\hbox{C}^{n(\gamma-1)-1\over(\gamma-1)}_{n}
{\widetilde H}_n\cr
}\eqno(24)
$$
It is interesting to note that the roots of the radical in (22) give the Riemann invariants of the problem, namely,
$$
-\lambda_{\pm}=u\pm{2\,v^{\gamma-1\over2}\over\gamma-1}
$$
and that both the polytropic gas and the nonlinear elastic medium with a polytropic gas potential share the same Riemann invariants.

The generating functions satisfy various identities. It is easy to check from the definitions in (22) that ($\sqrt{\hskip 1.0truecm}=\left\{\left({u+\lambda\over2}\right)^2-{v^{\gamma-1}\over(\gamma-1)^2}\right\}^{1\over2}$)
$$
\displaylines{
\hfill\eqalign{
{\partial{\chi}\over\partial u}=&{1\over\gamma-1}\,
{1\over2\sqrt{\hskip 1.0truecm}}\chi\cr
{\partial{\chi}\over\partial v}=&{v^{-1}\over2\sqrt{\hskip 1.0truecm}}\,
\left(\sqrt{\hskip 1.0truecm}-{u+\lambda\over2}\right)\chi\cr
}\hfill
\eqalign{
{\partial{\widetilde \chi}\over\partial u}=&-{1\over\gamma-1}\,
{1\over2\sqrt{\hskip 1.0truecm}}{\widetilde\chi}\cr
{\partial{\widetilde \chi}\over\partial v}=&{v^{-1}\over2\sqrt{\hskip 1.0truecm}}\,\left(\sqrt{\hskip 1.0truecm}+{u+\lambda\over2}\right){\widetilde \chi}\cr
}\hfill(25)}
$$ 
It follows trivially from these that
$$
\displaylines{
\hfill
\eqalign{
v^{\gamma-2}{\partial\chi\over\partial u}=&-{\gamma-1\over\gamma}\,
\lambda^{-1}\,
{\partial\chi^\gamma\over\partial v}\cr
v{\partial\chi\over\partial v}-\chi=&-{\gamma-1\over\gamma}\,\lambda^{-1}\,
{\partial\chi^\gamma\over\partial u}\cr}
\hfill
\eqalign{
v^{\gamma-2}{\partial{\widetilde \chi}\over\partial u}=&{\gamma-1\over\gamma}
\lambda^{-(\gamma-1)^2}\,
{\partial{\widetilde \chi}^\gamma\over\partial v}\cr
v{\partial{\widetilde \chi}\over\partial v}-{\widetilde \chi}=
&{\gamma-1\over\gamma}\lambda^{-(\gamma-1)^2}\,
{\partial{\widetilde \chi}^\gamma\over\partial u}\cr
}
\hfill(26)
}
$$
The consistency of these leads to
$$
{\partial^2\chi\over\partial v^2}=v^{\gamma-3}
{\partial^2\chi\over\partial u^2},\,\,\,\,\,\,\,
{\partial^2{\widetilde \chi}\over\partial v^2}=v^{\gamma-3}
{\partial^2{\widetilde \chi}\over\partial u^2}
\eqno(27)
$$
which is indeed in the form (6). We can now prove that the conserved charges are in involution  with respect to the Hamiltonian structure in (1) in a simple manner [4]. By definition
$$
\eqalign{
\left\{\chi(\lambda),\chi(\lambda')\right\}=&\int dx\,
\left({\partial\chi(\lambda)\over\partial u}{\partial\ \over\partial x}
{\partial\chi(\lambda')\over\partial v}+
{\partial\chi(\lambda)\over\partial v}{\partial\ \over\partial x}
{\partial\chi(\lambda')\over\partial u}
\right)\cr
=&\int dx\,\left(Fu_x+Gv_x\right)\cr
}\eqno(28)
$$
where
$$
\eqalign{
F\equiv&{\partial\chi(\lambda)\over\partial u}\,
{\partial^2\chi(\lambda')\over\partial u\partial v}+
{\partial\chi(\lambda)\over\partial v}\,
{\partial^2\chi(\lambda')\over\partial u^2}\cr
G\equiv&{\partial\chi(\lambda)\over\partial v}\,
{\partial^2\chi(\lambda')\over\partial u\partial v}+
{\partial\chi(\lambda)\over\partial u}\,
{\partial^2\chi(\lambda')\over\partial v^2}\cr
}\eqno(29)
$$
Using (27) we can show that
$$
{\partial F\over\partial v}-{\partial G\over\partial u}=0\eqno(30)
$$
which implies that
$$
F={\partial K\over\partial u},\,\,\,\,G={\partial K\over\partial v}
$$
Consequently, we obtain
$$
\left\{\chi(\lambda),\chi(\lambda')\right\}=\int dx\,
\left({\partial K\over\partial u}u_x+{\partial K\over\partial v}v_x\right)=\int dx{\partial K\over\partial x}=0
\eqno(31)
$$
Similarly, one can show
$$
\left\{{\widetilde \chi}(\lambda),{\widetilde \chi}(\lambda')\right\}=
\left\{\chi(\lambda),{\widetilde \chi}(\lambda')\right\}=0\eqno(32)
$$
Since the generating functions are in involution for arbitrary parameters $\lambda$ and $\lambda'$, it follows that all the conserved charges, ${\cal H}_n$ and ${\widetilde {\cal H}}_n$, must also be in involution with one another.

The generating functions, $\chi$ and $\widetilde \chi$, contain all the information about the system. One can easily study the symmetries, the recursion relation between the conserved quantities as well as the higher order Hamiltonian structures from these. Here we give the simple, but nontrivial example of how the recursion relations between the conserved charges as well as the second Hamiltonian structure for the system can be derived from the generating function. Let
$$
{\partial\ \over\partial t}={1\over(\gamma-1)^{2\over\gamma-1}}
\sum_{n=0}^\infty\lambda^{-n}{\partial\ \over\partial t_n}\eqno(33)
$$
Then, we can write the entire hierarchy of polytropic gas equations in the compact form
$$
{\partial\ \over\partial t}\pmatrix{
u\cr
\noalign{\vskip .5truecm}%
v}={\cal D}_0
\pmatrix{
{\partial{\widetilde\chi}\over\partial u}\cr
\noalign{\vskip .2truecm}%
{\partial{\widetilde\chi}\over\partial v}}=
\pmatrix{0&\partial\cr
\noalign{\vskip .5truecm}%
\partial&0}
\pmatrix{
{\partial{\widetilde\chi}\over\partial u}\cr
\noalign{\vskip .2truecm}%
{\partial{\widetilde\chi}\over\partial v}}\eqno(34)
$$
From Eq. (25), we can write
$$
\eqalign{
(\gamma-1)(u+\lambda){\partial{\widetilde\chi}\over\partial u}+2v
{\partial{\widetilde\chi}\over\partial v}-{\widetilde \chi}&=0\cr
(\gamma-1)\left(2\sqrt{\hskip 1.0truecm}-(u+\lambda)\right){\partial{\widetilde\chi}\over\partial v}-2v^{\gamma-2}
{\partial{\widetilde\chi}\over\partial u}&=0\cr
}\eqno(35)
$$
Taking the derivative of these with respect to $x$, we obtain
$$
\eqalign{
-(\gamma-1)\lambda\,\partial_x{\partial{\widetilde\chi}\over\partial u}=&
(\gamma-2)u_x{\partial{\widetilde\chi}\over\partial v}+v_x{\partial{\widetilde\chi}\over\partial v}+(\gamma-1)u\,\partial_x {\partial{\widetilde\chi}\over\partial u}+2v\,\partial _x{\partial{\widetilde\chi}\over\partial v}\cr
-(\gamma-1)\lambda\,\partial_x{\partial{\widetilde\chi}\over\partial v}=&
(\gamma-2)v^{\gamma-3}v_x{\partial{\widetilde\chi}\over\partial u}+2v^{\gamma-2}\partial_x{\partial{\widetilde\chi}\over\partial u}+u_x 
{\partial{\widetilde\chi}\over\partial v}+(\gamma-1)u\,\partial_x{\partial{\widetilde\chi}\over\partial v}\cr
}\eqno(36)
$$
Restricting to specific powers of $\lambda$, these, then define a recursion relation between the conserved densities (with suitable normalization), namely,
$$
\eqalign{
\partial_x{\partial{\widetilde H}_{n+1}\over\partial u}=&
(\gamma-2)u_x{\partial{\widetilde H}_n\over\partial v}+v_x{\partial{\widetilde H}_n\over\partial v}+(\gamma-1)u\,\partial_x {\partial{\widetilde H}_n\over\partial u}+2v\,\partial _x{\partial{\widetilde H}_n\over\partial v}\cr
\partial_x{\partial{\widetilde H}_{n+1}\over\partial v}=&
(\gamma-2)v^{\gamma-3}v_x{\partial{\widetilde H}_n\over\partial u}+2v^{\gamma-2}\partial_x{\partial{\widetilde H}_n\over\partial u}+u_x 
{\partial{\widetilde H}_n\over\partial v}+(\gamma-1)u\,\partial_x{\partial{\widetilde H}_n\over\partial v}\cr
}\eqno(37)
$$

We note here that the recursion relation in (36) can equivalently be written as
$$
-(\gamma-1)\,\lambda\,{\cal D}_0
\pmatrix{
{\partial{\widetilde\chi}\over\partial u}\cr
\noalign{\vskip .2truecm}%
{\partial{\widetilde\chi}\over\partial v}}=
{\cal D}_1
\pmatrix{
{\partial{\widetilde\chi}\over\partial u}\cr
\noalign{\vskip .2truecm}%
{\partial{\widetilde\chi}\over\partial v}}\eqno(38)
$$
where
$$
{\cal D}_1=\pmatrix{
\partial v^{\gamma-2}+v^{\gamma-2}\partial&\partial u+(\gamma-2)u\partial\cr
(\gamma-2)\partial u+u\partial &\partial v+v\partial
}\eqno(39)
$$
Using this in the dynamical equation, we note that we can write
$$
{\partial\ \over\partial t}\pmatrix{
u\cr
\noalign{\vskip .5truecm}%
v}={\cal D}_0
\pmatrix{
{\partial{\widetilde\chi}\over\partial u}\cr
\noalign{\vskip .2truecm}%
{\partial{\widetilde\chi}\over\partial v}}=
-{\lambda^{-1}\over\gamma-1}{\cal D}_1
\pmatrix{
{\partial{\widetilde\chi}\over\partial u}\cr
\noalign{\vskip .2truecm}%
{\partial{\widetilde\chi}\over\partial v}}\eqno(40)
$$
which shows that the hierarchy of equations is bi-Hamiltonian with ${\cal D}_1$ representing the second Hamiltonian structure obtained in [5].

Finally, we note here that the system of polytropic gas equations has a natural zero curvature representation in this framework [3]. Let
$$
{\cal A}_1=\pmatrix{p-u & 1\cr v & -p},\,\,\,\,
{\cal A}_0=\pmatrix{{1\over4}u^2+{1\over2}{v^{\gamma-1}\over\gamma-1} &-{1\over2}u\cr -{1\over2}uv & -{1\over2}v\cr}\eqno(41)
$$
then, it is easy to check that
$$
\partial_t{\cal A}_1-\partial_x{\cal A}_0-\{{\cal A}_0,{\cal A}_1\}=0\eqno(42)
$$
gives the equations for the polytropic gas in $(10)$.
\bigskip
\leftline{\bf Acknowledgments}
\medskip
 
We acknowledge with pleasure the impetus provided by Baianinha for the completion of this work. A. D. would like to thank the members of the Departamento de F\'\i sica at UFSC for hospitality during the period when this work was done.
J.C.B. was supported by CNPq, Brazil. A.D. was supported in part by the U.S. 
Department of Energy Grant No. DE-FG-02-91ER40685 and by NSF-INT-9602559. 

\vskip 2.0truecm


\leftline{\bf References}
\bigskip

\item {1.} Yu. I. Manin, J. Sov. Math. {\bf 11}, 1 (1979).

\item{2.} B. A. Dubrovin and S. P. Novikov, Russian Math. Surveys  {\bf 44}, 
35 (1989).

\item{3.} V. E. Zakharov, Funct. Anal. Appl. {\bf 14}, 89 (1980).

\item{4.} J. Cavalcante and H. P. McKean, Physica {\bf 4D}, 253 (1982).

\item{5.} P. J. Olver and Y. Nutku, J. Math. Phys. {\bf 29}, 1610 (1988).

\item{6.} G. B. Whitham, ``Linear and Nonlinear Waves'' (Wiley, New
York, 1974).

\item{7.} J. C. Brunelli, Rev. Math. Phys. {\bf 8}, 1041 (1996).
								 
\item{8.} B. Enriquez, A. Yu. Orlov and V. N. Rubtsov, Inverse Problems 
{\bf 12}, 241 (1996).

\end